\documentclass{aa}  

\usepackage{graphicx}
\usepackage{txfonts}
\usepackage{hyperref}
\newcommand*{\mj}{\ensuremath{M_{\mathrm{J}}}}
\usepackage[utf8]{inputenc}
\usepackage{multirow}

\makeatletter
\renewcommand*\aa@pageof{, page \thepage{} of \pageref*{LastPage}}
\makeatother

\def\xHyphenate#1#2\wholeString {\if#1$%
    \else\transform{#1}%
    \takeTheRest#2\ofTheString\fi}

\def\takeTheRest#1\ofTheString\fi
{\fi \xHyphenate#1\wholeString}

\def\transform#1{\url{#1}\hskip 0pt plus 1pt}

\makeatletter
\def\xhref#1#2{\def\tmp{#1}\xxhref#2 \^ }
\def\xxhref#1 {%
  \ifx\^#1\expandafter\@gobble\else\expandafter\@firstofone\fi
  {\mbox{\href{\tmp}{#1}} \xxhref}}
\makeatother

\begin{document}

   \title{MADYS: the Manifold Age Determination for Young Stars\thanks{Available at \href{https://github.com/vsquicciarini/madys}{https://github.com/vsquicciarini/madys}.}}
   \subtitle{I. Isochronal age estimates and model comparison}

   \author{V. Squicciarini
          \inst{1,2}
          \and
          M. Bonavita\inst{1,3}
          }

   \institute{Department of Physics and Astronomy ‘Galileo Galilei’, University of Padova; Vicolo dell’Osservatorio 3, I-35122 Padova, Italy\\
              \email{vito.squicciarini@inaf.it}
         \and
             INAF – Osservatorio Astronomico di Padova; Vicolo dell’Osservatorio 5, I-35122 Padova, Italy
        \and
            School of Physical Sciences, Faculty of Science, Technology, Engineering and Mathematics, The Open University; Walton Hall, Milton Keynes, MK7 6AA, UK
             }

   \date{Received --; accepted --}

  \abstract
  {The unrivalled astrometric and photometric capabilities of the Gaia mission have given new impetus to the study of young stars: both from an environmental perspective, as members of comoving star-forming regions, and from an individual perspective, as targets amenable to planet-hunting direct-imaging observations.}
  {In view of the large availability of theoretical evolutionary models, both fields would benefit from a unified framework that allows a straightforward comparison of physical parameters obtained by different stellar and substellar models.}
  {To this aim, we developed the Manifold Age Determination for Young Stars (\textsc{madys}), a flexible Python tool for the age and mass determination of young stellar and substellar objects. In this first release, \textsc{madys} automatically retrieves and crossmatches photometry from several catalogs, estimates interstellar extinction, and derives age and mass estimates for individual objects through isochronal fitting.}
  {Harmonizing the heterogeneity of publicly available isochrone grids, the tool allows one to choose amongst 17 models, many of which with customizable astrophysical parameters, for a total of $\sim 110$ isochrone grids. Several dedicated plotting functions are provided to allow for an intuitive visual perception of the numerical output.}
  {After extensive testing, we have made the tool publicly available. Here, we demonstrate the capabilities of \textsc{madys}, summarizing previously published results as well providing several new examples.}


   \maketitle
%

\section{Introduction}

The advent of Gaia \citep{2016A&A...595A...1G} has brought our view of the Galaxy to its grandest level of sharpness, opening up tremendous possibilities that stretch from the direct detection of exoplanets \citep[e.g.,][]{2022MNRAS.513.5588B} to the realm of galactic archaeology \citep[e.g.,][]{2018Natur.563...85H}. The availability of precise photometric and distance measurements for $\sim 2 \cdot 10^9$ stars enabled by Gaia Data Release 3 \citep[Gaia DR3;][]{2022arXiv220800211G} has paved the way for precise large-scale measurements of stellar luminosity and effective temperature, which in turn allow one to discern exquisite features inside theoretical Hertzsprung-Russell diagrams (HRDs) or their observational counterpart, color-magnitude diagrams (CMDs) \citep{2018A&A...616A..10G}.

The position of a star in a CMD, corrected for distance and interstellar extinction, can in principle be compared with theoretical grids derived from stellar evolution models to derive several astrophysical parameters, including mass, age, and metallicity. However, this widely used technique, known as "isochrone fitting," is weakened by a strong degeneracy between age and metallicity \citep{2019A&A...622A..27H}. Even when metallicity is independently assessed, obtaining precise ages is an extremely challenging task for main-sequence (MS) stars \citep{2010ARA&A..48..581S}: the reason lies in the feeble variation of luminosity and effective temperature over the main-sequence lifetime, keeping the isochrones very close to one another in the HRDs.

Substantial efforts have thus been undertaken to explore complementary age-dating techniques\footnote{Each one focusing on specific phases of stellar evolution and/or mass ranges; see \citet{2016EAS....80..115B} for a review.} such as gyrochronology \citep{2007ApJ...669.1167B,2014ApJ...780..159E}, chromospheric activity \citep{2008ApJ...687.1264M,2021ApJ...908..207Z}, lithium depletion \citep{2014prpl.conf..219S}, and, very promisingly, asteroseismology \citep[e.g.,][]{2014MNRAS.445.2758R,2016MNRAS.456.3655M,2021MNRAS.502.1947M}. Over the last few years, first steps have been taken in combining infrared photometry from the Two Micron All-Sky Survey \citep[2MASS;][]{2006AJ....131.1163S} and AllWISE \citep{2014yCat.2328....0C} with spectroscopic constraints on effective temperature and surface gravity to derive stellar parameters for millions of main-sequence and post-main sequence stars consistently \citep[e.g.,][]{2020AJ....160...83S,2020arXiv201209690M,2022A&A...662A.125F}. The conjunction of isochrone fitting and gyrochronology is the idea behind \textsc{stardate} \citep{2019AJ....158..173A}, which shows that the integration of rotational periods into the classical isochrone-based framework can lead to a gain of up to a factor three in the accuracy of the age for main-sequence FGK stars.

In addition to this, the exquisite precision of Gaia's proper motion measurements has been enormously beneficial to the study of stellar moving groups, associations and clusters, leading to the compilation of large catalogs of confirmed members for known regions \citep[e.g.,][]{2018A&A...618A..93C,2018ApJ...856...23G,2020AJ....160...44L} and even to the discovery of new ones \citep[e.g.,][]{2019A&A...624A.126C}. A complete census of star-forming regions is, in turn, the first step toward resolving kinematic substructures within them and connecting these structures with star formation history \citep[e.g.,][]{2019ApJ...870...32K,2019A&A...626A..17C,2022ApJ...931..156P}. Again, the need for differential age estimates within star-forming regions implies precise individual age determinations; the picture is further complicated by the fact that, at such young ages ($t \lesssim 100$ Myr), a large fraction of the stars is still in the pre-main sequence (PMS) phase. 

In principle, the dependence of luminosity on age during the PMS is steep enough to allow a simultaneous determination of age and mass; indeed, the spread between age predictions by different models -- due to factors like initial helium abundance, metallicity, mixing length, convective core overshooting \citep{2016AN....337..819P}-- is generally acceptable ($\sim 20\%$) for F-G stars \citep{2010ARA&A..48..581S}. However, the accuracy rapidly degrades at later spectral types, so that the inter-model spread can be as high as a factor of 4-5 for late-K and M stars \citep{2010ARA&A..48..581S} as a consequence of rotation, activity and magnetic fields \citep{2016IAUS..314...79F}.

Not surprisingly, the problem continues to exacerbate below the hydrogen-burning limit ($\sim 0.08 M_\odot$) which separates stars from brown dwarfs \citep{2012ApJ...745..174S}. Thanks to the development of high-contrast imaging facilities, young luminous brown dwarfs and giant exoplanets are being increasingly found both in isolation \citep{2022NatAs...6...89M} and as companions to members of moving groups and associations \citep[e.g.,][]{2021A&A...651A..72V}. In this case, a simultaneous isochronal estimate of age and mass is no more feasible, and independent constraints are needed to lift the degeneracy between age and mass. Given the importance of the derived properties of these young substellar objects, to study the low-mass end of the stellar initial mass function (IMF) on the one hand \citep[e.g.,][]{2019ApJS..240...19K}, and exoplanet demographics on the other hand \citep[e.g.,][]{2019AJ....158...13N}, it becomes a crucial task to compare the different predictions done by different models. In fortunate cases, mass estimates can be compared to model-independent dynamical masses, possibly disentangling among formation mechanisms \citep[e.g.,][]{2014MNRAS.437.1378M,2021AJ....161..179B}.

With an eye on the study of star-forming regions and the other on directly imaged substellar objects, we developed the Manifold Age Determination of Young Stars (\textsc{madys}): a tool that allows a robust inference of the properties of stellar (substellar) objects based on stellar (substellar) evolution models, and a straightforward comparison between different suites of models. 
In this first paper we present the first, fully documented, public release of \textsc{madys}\footnote{Our tool is available at \url{https://github.com/vsquicciarini/madys}.}, a preliminary version of which was already used in several publications \citep{2021A&A...646A.164J,2021MNRAS.507.1381S,2022A&A...663A.144B,2022A&A...664A...9S}, and the underlying algorithm for isochronal age and mass determination. We defer to a future paper the implementation of an indirect method for age determination in star-forming regions, based on empirical kinematic properties \citep[see ][]{2021MNRAS.507.1381S}.

This paper is structured as follows. Section~\ref{sec:introd_madys} introduces the key concepts behind \textsc{madys}, with a special focus on data retrieval and extinction estimates; the algorithm for age and mass determination is presented in Section~\ref{sec:agemass}; a few applications are then provided in Section~\ref{sec:applications}. In Section~\ref{sec:discussion}, we discuss the strengths and the limitations of the tool, and anticipate its future developments. Finally, in Section~\ref{sec:conclusions} we give a short summary of our results and the main conclusions.

\section{Introducing MADYS}
\label{sec:introd_madys}

The Manifold Age Determination of Young Stars (\textsc{madys}), written in Python, is a tool that allows to derive the properties of arbitrarily large lists of young objects by comparison with several published evolution models. \textsc{madys} combines extensive cross-match capabilities, a careful assessment of photometric data, the ability to estimate interstellar extinction in hundreds of photometric bands, and the homogenization of a large collection of stellar and substellar evolutionary models. The derivation of ages and masses can be performed under several configurations depending on the science case, allowing for the presence of multimodal posterior distributions.

The tool can be used in two different modes, depending on the kind of input provided: either a list of object names ({\it mode 1}) or a table containing object names, photometry and, optionally, astrometry ({\it mode 2}).

\noindent Several examples of code execution are provided in a dedicated Jupyter Notebook within the GitHub repository\footnote{\url{https://github.com/vsquicciarini/madys/blob/main/examples/MADYS_examples.ipynb}}.

Generally speaking, the execution of the program triggers the following chain of operations: after retrieving --but only in mode 1, see Section~\ref{sec:data_retrieval}-- photometric and astrometric data, reliable photometry is identified (Section~\ref{sec:phot_sel}); then, the estimation of interstellar extinction in all the bands of interest is performed (Section~\ref{sec:ext}), resulting in a final database to be used for age and mass determination (Section~\ref{sec:age_mass}).

The estimation of physical parameters is not done during initialization, but rather by calling a dedicated method that acts upon the database: in this way, it is possible to inspect data, to carefully decide the (sets of) theoretical model(s) suitable to the science case (Section~\ref{sec:load_iso}), and to repeat multiple times the analysis of the same database.

\subsection{Data retrieval}
\label{sec:data_retrieval}

\subsubsection{Mode 1}
\label{sec:data_retrieval1}

Building on the capabilities of \texttt{astroquery}\footnote{\href{https://astroquery.readthedocs.io/en/latest/}{https://astroquery.readthedocs.io/en/latest/}} and \textsc{tap}\footnote{\href{https://github.com/mfouesneau/tap}{https://github.com/mfouesneau/tap}.} to handle existing cross-matches between Gaia and other catalogs \citep{2019A&A...621A.144M}, \textsc{madys} queries the Gaia Archive to return a single catalog containing astrometric, kinematic and photometric information from Gaia Data Release 2 \citep[Gaia DR2;][]{2018A&A...616A...1G}, Gaia DR3 \citep{2022arXiv220800211G} and 2MASS \citep{2006AJ....131.1163S}. Optionally, AllWISE \citep{2014yCat.2328....0C}, Pan-STARRS  Data Release 1 \citep{2016arXiv161205560C}, and the Sloan Digital Sky Survey Data Release 3 \citep[SDSS DR13;][]{2017ApJS..233...25A} can be added to the query as well\footnote{A few additional catalogs (listed in \xhref{https://gea.esac.esa.int/archive/documentation/GEDR3/Catalogue_consolidation/chap_crossmatch/sec_crossmatch_externalCat/}{https://gea.esac.esa.int/archive/documentation/GEDR3/Catalogue \_consolidation/chap\_crossmatch/sec\_crossmatch\_externalCat/}) might be incorporated in future versions of the program, if considered useful by the community.}.

Although it is recommended to use Gaia DR2 or Gaia DR3 IDs, it is also possible to use other naming conventions. In the latter case, input names are converted into their Gaia DR2 counterparts through a query of the SIMBAD database \citep{2000A&AS..143....9W}.

The results of the query are provided as a table whose $i$-th row corresponds to the $i$-th star of the initial list of objects to allow an unambiguous pairing of input and output data. In other words, the query algorithm of \textsc{madys} is able to handle the cases in which more than one Gaia DR3 source is associated to the same Gaia DR2 source (or vice versa), selecting as best-mach the source with Gaia DR2 ID = Gaia DR3 ID or, if missing, the source in the VizieR catalog \citep{2000A&AS..143...23O} having the closest $G$, $G_{BP}$ and $G_{RP}$ photometry in a suitable region accounting for the astrometric motion of the source over the two epochs \footnote{In the latter case, the source $X$ is considered a cross-match to source $0$ only if $|G_0-G_X|<0.2$ mag and $|G_{BP,0}-G_{BP,X}|<0.2$ mag and $|G_{RP,0}-G_{BP,X}|<0.2$ mag.}. This stratagem is able to find a cross-match for some high-proper motion stars which are not paired by the SIMBAD database. Likewise, a small ($\sim 0.5 \%$) fraction of missing 2MASS matches is recovered by indirectly exploiting the AllWISE-2MASS cross-match, or --if explicitly required-- by directly inspecting the SIMBAD database.

The typical speed of the query is about $\sim 100-150$ stars s$^{-1}$, meaning that a list of 1000 objects is fully recovered within few seconds. Large lists of objects ($\sim 10^4-10^5$) are handled efficiently by splitting the query into smaller chunks and later reassembling the results in the original order.

In any case, the resulting database always comprises data from Gaia DR2 and Gaia DR3. Parallaxes from Gaia DR3\footnote{If the parallax for a source is present in DR2 but not in DR3, values from Gaia DR2 are used.} and extinctions (Section~\ref{sec:ext}) are combined with apparent photometry to get absolute magnitudes. Quality flags from selected surveys are retained with the aim of identifying reliable photometric data (Section~\ref{sec:phot_sel}).

\subsubsection{Mode 2}
\label{sec:data_retrieval2}
In mode 2, a table containing full information needed for age and mass determination is provided. This mode is thought for objects that are not present in Gaia, such as self-luminous direct-imaged exoplanets and brown dwarfs.

Minimum requirements, in this case, consist of a column of object names and a column with magnitudes. If parallaxes are provided, input magnitudes are considered as apparent; otherwise, they are considered as absolute. By providing two columns with equatorial or galactic coordinates, it is possible for the program to evaluate interstellar extinction in the direction of the object(s) and to take it into account (see Section~\ref{sec:ext}). Alternatively, values for the E(B-V) color excess can be manually provided.

More than 250 photometric filters are available in this mode, meaning that there is at least one theoretical model which they can be compared to. The full list of filters --including, for example, the full suite of the James Webb Space Telescope \citep[JWST,][]{2006SSRv..123..485G} filters-- can be found in the documentation.

\subsection{Selection of appropriate photometry values}
\label{sec:phot_sel}

This Section describes the conditions for a photometric measurement to be retained in the final database. By default, \textsc{madys}' mode 1 collects photometric measurements from Gaia DR2/DR3 ($G$, $G_{BP}$, $G_{RP}$) and 2MASS ($J$, $H$, $K_s$). Gaia DR3 $G$ magnitudes are corrected by adopting the prescriptions by \citet{2021A&A...649A...3R}. As regards $G_{BP}$ and $G_{RP}$, which are known to be intrinsically much more sensitive than $G$ to contamination from nearby sources or from the background \citep{2018A&A...616A...4E}, the phot\_bp\_rp\_excess\_factor $C$ is used as a proxy to evaluate the quality of photometric measurements. Following \citet{2021A&A...649A...3R}, a color-independent corrected $BP/RP$ excess factor $C^*$ was defined for both Gaia DR2 and Gaia DR3:

\begin{equation}
\label{eq:excess_f}
C^* = C + k_0+k_1\Delta G+k_2 \Delta G^2+k_3 \Delta G^3+k_4 G
\end{equation}
where $\Delta G=(G_{BP}-G_{RP})$.

The corrected $BP/RP$ excess factor has an expected value of 0 for well-behaved sources at all magnitudes but, when considering subsamples of stars with similar brightness, it tends to widen out for fainter $G$; a varying standard deviation $\sigma(G)$ can be defined \citep{2021A&A...649A...3R} as follows:

\begin{equation}
\label{eq:excess_f_sigma}
\sigma_{C^*}(G)=c_0+c_1 \cdot G^m.
\end{equation}

\noindent Values for the constants for Eq.~\ref{eq:excess_f}-~\ref{eq:excess_f_sigma} are taken from \citet{2021A&A...649A...3R} for DR3 and \citet{2021MNRAS.507.1381S} for DR2, and are provided in Table ~\ref{tab:excess_param}.

We exclude $G_{BP}$ and $G_{RP}$ magnitudes with a corrected excess factor larger, in absolute value, than 3 $\sigma_{C^*}(G)$. As mentioned above, a value of $C^*$ significantly different from zero might be due to blended Gaia transits or crowding effects; in addition to this, it can also be related to an over-correction of the background (if $C^*$<0) or to an anomalous SED (if $C^*$>0) characterized by strong emission lines in the wavelength window where the $G_{RP}$ transmissivity is larger than the $G$ transmittivity. This latter case can occur, for instance, for a source located in a HII region \citep[see discussion in ][]{2021A&A...649A...3R}.

\begin{table}
  \centering
   \caption{Adopted values for Eq.~\ref{eq:excess_f}-\ref{eq:excess_f_sigma}.}
   \begin{tabular}{c|ccc} \hline
DR2 & $\Delta G < 0.5$ & $ \text{ } 0.5 \leq \Delta G < 3.5$ & $\Delta G \geq 3.5$ \\
            \hline
$k_0$ & -1.121221 & -1.1244509 & -0.9288966 \\
$k_1$ & +0.0505276 & +0.0288725 & -0.168552 \\
$k_2$ & -0.120531 & -0.0682774 & 0 \\
$k_3$ & 0 & 0.00795258 & 0 \\
$k_4$ & -0.00555279&-0.00555279 & -0.00555279 \\ \hline
$c_0$ & \multicolumn{3}{c}{0.004} \\
$c_1$ & \multicolumn{3}{c}{$8\cdot 10^{-12}$} \\
$m$ & \multicolumn{3}{c}{7.55} \\
            \hline \hline
DR3 & $\Delta G < 0.5$ & $ \text{ } 0.5 \leq \Delta G < 4$ & $\Delta G \geq 4$ \\
            \hline
$k_0$ & 1.154360 & 1.162004 & 1.057572 \\
$k_1$ & 0.033772 & 0.011464 & 0.140537 \\
$k_2$ & 0.032277 & 0.049255 & 0 \\
$k_3$ & 0 & -0.005879 & 0 \\
$k_4$ & 0 & 0 & 0 \\ \hline
$c_0$ & \multicolumn{3}{c}{0.0059898} \\
$c_1$ & \multicolumn{3}{c}{$8.817481\cdot 10^{-12}$} \\
$m$ & \multicolumn{3}{c}{7.618399} \\
            \hline
   \end{tabular}
   \label{tab:excess_param}
\end{table}
\noindent

From 2MASS and AllWISE, only sources with photometric flag ph\_qual='A' are kept. If needed, a different value for the worst quality flag still considered reliable can be selected.

\subsection{Interstellar extinction}
\label{sec:ext}

The estimate of extinction (reddening) in a given band (color) is performed by integrating along the line of sight a suitable 3D extinction map. The integration algorithm draws a line from the position of the Sun toward that of the star of interest; the value of each pixel crossed by the line is weighted according to the portion of the total distance spent by the line in the pixel itself. This method ensures a rapid and precise evaluation of the integral, allowing 10000 stars to be handled in $\sim 1$ s under typical PC performances.

\noindent Two extinction maps can be selected:
\begin{itemize}
\item the STILISM 3D extinction map by \citet{2019A&A...625A.135L}: a Sun-centered (6000x6000x800) pc grid, with step equal to 5 pc;
\item the Galactic extinction catalog by \citet{2020A&A...639A.138L}: a Sun-centered (740x740x540) pc grid with step equal to 1 pc.
\end{itemize}

Since the file with the selected map must be present in the local path where \textsc{madys} has been installed, the errors on the derived estimates --which would require the download of additional files-- are currently not returned by the program.

Coordinate transformations from the equatorial or galactic frame to the right-handed galactocentric frame (i.e., a Cartesian galactic frame) is performed by means of the astropy package\footnote{Default parameters from the "pre-v4.0" are used: galcen\_distance=8.3 kpc, galcen\_v\_sun=(11.1, 232.24, 7.25) km s$^{-1}$, z\_sun=27.0 pc, roll=0.0 deg. See \xhref{https://docs.astropy.org/en/stable/api/astropy.coordinates.Galactocentric.html}{https://docs.astropy.org/en/stable/api/astropy.coordinates. Galactocentric.html} for details.} \citep{2013A&A...558A..33A}.

\begin{figure*}
	\includegraphics[width=\hsize]{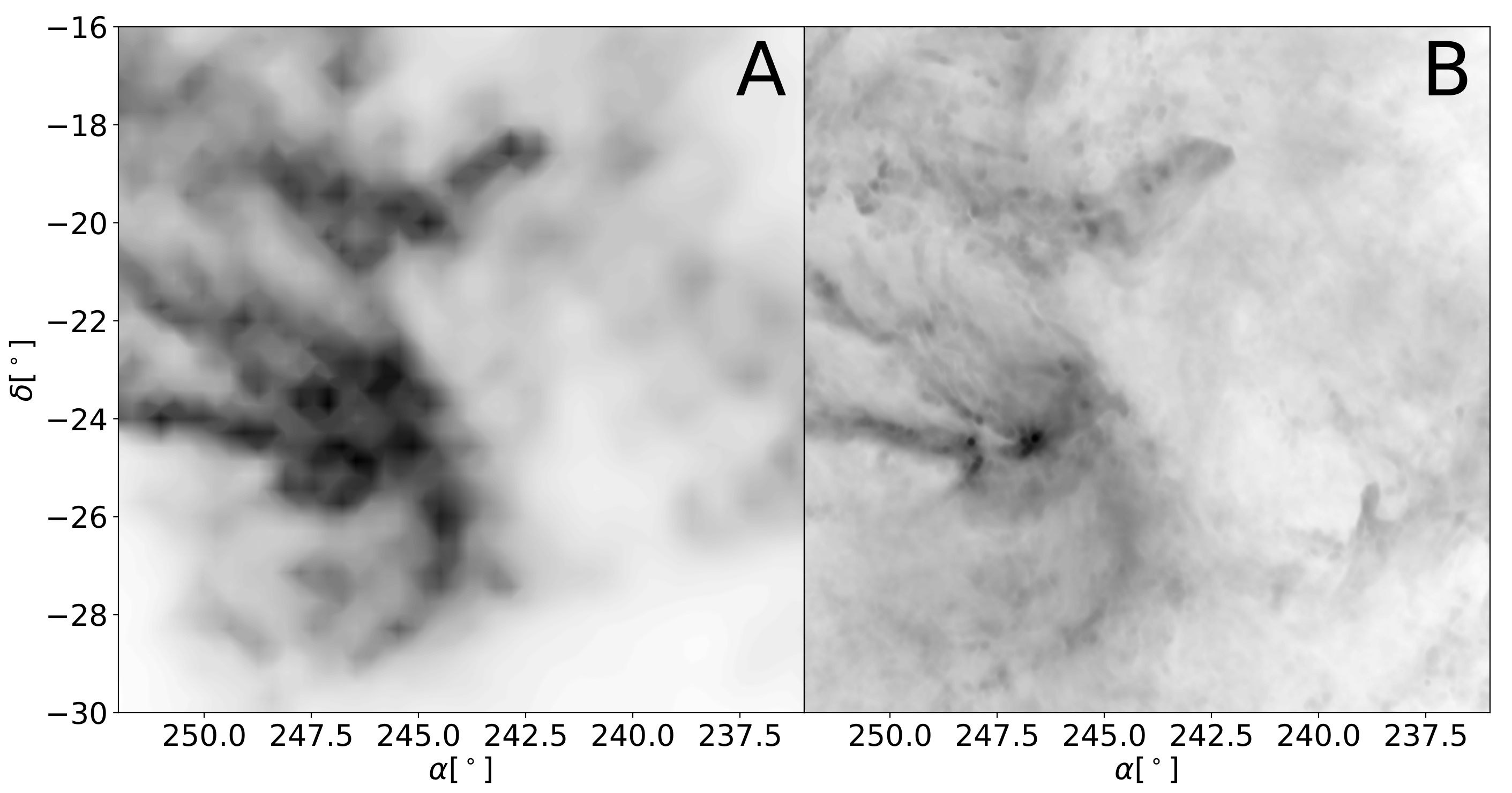}
    \caption{Integrated extinction toward Upper Scorpius. (A) G-band extinction map produced by \textsc{madys} by integrating the 3D map by \citet{2020A&A...639A.138L} up to a distance $d=160$ pc. (B) Intensity Stokes map at $350~\mu$m produced by the Planck satellite.}
    \label{fig:US_extinction}
\end{figure*}

As a test for the accuracy of the algorithm, we provide in Figure~\ref{fig:US_extinction} a visual comparison between the integrated absorption in the Upper Scorpius region \citep[already used in ][]{2021MNRAS.507.1381S} obtained through the map by \citet{2020A&A...639A.138L} and the intensity Stokes map returned by the Plank satellite \citep{2020A&A...641A...1P} in the far infrared ($\nu=857$ GHz $= 350~\mu$m). Given the large galactic latitude of the region ($l \in [10^\circ,30^\circ]$), we expect the large majority of the integrated intensity in the Planck image to be produced by the association, with only a negligible background contribution. Indeed, the agreement between the two images is excellent.

\begin{figure}
	\includegraphics[width=\hsize]{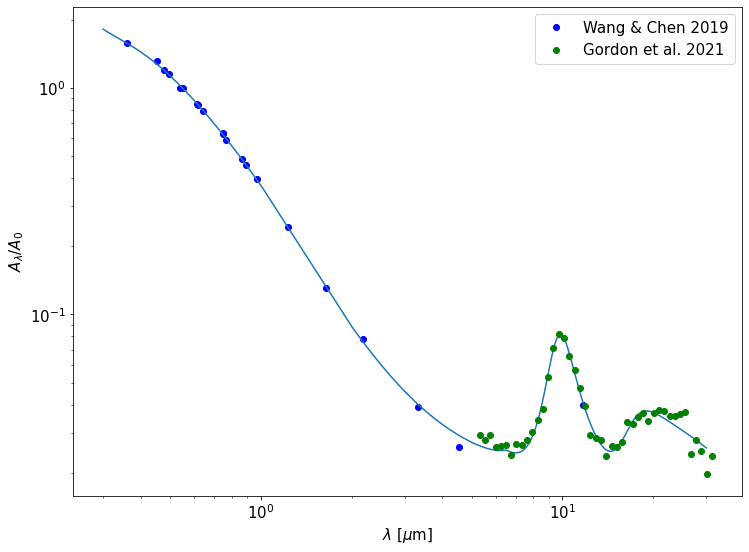}
    \caption{Adopted extinction law (solid line). Blue dots represent absorption coefficients in the visible and NIR \citep[][Table 3]{2019ApJ...877..116W}, green dots in the MIR \citep[][Table 8]{2021ApJ...916...33G}.}
    \label{fig:ext_law}
\end{figure}

The conversion between extinction and reddening is mediated by a total-to-selective absorption ratio $R=3.16$ \citep{2019ApJ...877..116W}. The extinction law is obtained by combining the extinction law by \citet{2019ApJ...877..116W} in the range $[0.3,2]$ $\mu$m and the diffuse average extinction by \citet{2021ApJ...916...33G} in the range $[6.5,40]$ $\mu$m; a linear combination of the two is used in the intermediate range $[2, 6.5]$ $\mu$m (Figure~\ref{fig:ext_law}):

\begin{equation} \label{eq:ir_law}
\frac{A_\lambda}{A_V} =
  \begin{cases}
    \displaystyle\sum_{i=0}^{7} b_i~\xi^i & ~\lambda \in [0.3,1]~\mu m \\
    h_2 \lambda^{\beta_2} \equiv f_2(\lambda) & ~\lambda \in [1,2]~\mu m \\
    [1-q(\lambda)] f_2(\lambda) + q(\lambda) f_4(\lambda) & ~\lambda \in [2,6.5]~\mu m \\
    h_4 \lambda^{\beta_4}+S_1 D_1(\lambda)+S_2 D_2(\lambda) \equiv f_4(\lambda) &  ~\lambda \in [6.5,40]~\mu m \\
  \end{cases} ,
\end{equation}

\noindent where:
\begin{equation}
    \xi = \frac{1}{\lambda}-1.82~\mu \text{m}^{-1}
\end{equation}
\noindent and
\begin{equation}
    q(\lambda) = \frac{\lambda-2~\mu \text{m}}{6.5~\mu \text{m}-2~\mu \text{m}},
\end{equation}

\noindent while $D_1$ and $D_2$ are two modified Drude profiles, used to model the silicate absorption features at $\sim 10~\mu \text{m}$ and $\sim 20~\mu \text{m}$:

\vspace{-5mm}
\begin{equation}\label{eq:ir_D}
    D(\lambda) = \frac{(\gamma(\lambda)/\lambda_0)^2}{((\lambda/\lambda_0-\lambda_0/\lambda)^2+(\gamma(\lambda)/\lambda_0)^2)}.
\end{equation}
Finally, $\gamma(\lambda)$ is in turn given by:
\begin{equation}\label{eq:ir_gamma}
    \gamma(\lambda) = \frac{2 \gamma_0}{1+\exp(a_0 (\lambda-\lambda_0))}
\end{equation}

\noindent \citep{2021ApJ...916...33G}. We list in Table~\ref{tab:ir_param} all the coefficients from Eq.~\ref{eq:ir_law}-\ref{eq:ir_gamma}, where ($\gamma_1$, $\lambda_1$, $a_1$) and ($\gamma_2$, $\lambda_2$, $a_2$) indicate the coefficients for $D_1(\lambda)$ and $D_2(\lambda)$, respectively.

\begin{table}
  \centering
   \caption{Adopted values for the coefficients in Eq.~\ref{eq:ir_law}-\ref{eq:ir_gamma}.}
   \begin{tabular}{cccc} \hline
Name & Value & Name & Value \\
            \hline
$b_0$ & 1 & $h_4$ & 0.366 \\
$b_1$ & 0.7499 & $\beta_4$ & -1.48 \\
$b_2$ & -0.1086 & $S_1$ & 0.06893 \\
$b_3$ & -0.08909 & $S_2$ & 0.02684 \\
$b_4$ & 0.02905 & $\lambda_1$ & 9.865 \\
$b_5$ & 0.01069 & $\gamma_1$ & 2.507 \\
$b_6$ & 0.001707 & $a_1$ & -0.232 \\
$b_7$ & -0.001002 & $\lambda_2$ & 19.973 \\
$h_2$ & 0.3722 & $\gamma_2$ & 16.989 \\
$\beta_2$ & -2.07 & $a_2$ & -0.273 \\
            \hline
   \end{tabular}
   \label{tab:ir_param}
\end{table}

The adopted extinction law goes farther in the mid-infrared than widely used parametrizations, as those offered by the extinction package\footnote{\href{https://extinction.readthedocs.io/en/latest/}{ttps://extinction.readthedocs.io/en/latest/}}, delving into wavelength ranges amenable to forthcoming JWST observations. Individual extinction coefficients $A_\lambda$ are directly taken from Table 3 of \citet{2019ApJ...877..116W} whenever possible, or computed through Eq.~\ref{eq:ir_law} adopting as $\lambda$ the mean wavelength indicated by the SVO Filter Profile Service \citep{2012ivoa.rept.1015R,2020sea..confE.182R}.

We would like to highlight that in the youngest ($t \lesssim 5$ Myr) star-forming regions, owing to the uneven and fragmentary nature of dust structures, the spatial variation of extinction usually occurs on smaller scales than when sampled by the available 3D maps. This limitation of the program can be currently handled by manually providing a vector of individual extinction values at initialization; a future version of \textsc{MADYS} will enable the simultaneous fit of extinction and (sub)stellar parameters based on available photometry.

\section{Age and mass determination}
\label{sec:agemass}

\subsection{Loading isochrone tables}
\label{sec:load_iso}

As already mentioned, the determination of ages and masses in this first release of \textsc{madys} is performed via isochronal fitting, comparing the database obtained in Section~\ref{sec:phot_sel} to the selected set of isochrones. We refer to these estimates as "photometric" or "isochronal" estimates interchangeably.

Overcoming the multifarious conventions of isochrone tables found in the literature, \textsc{madys} employs an extended version of the \texttt{evolution} routine\footnote{\href{https://github.com/avigan/Python-utils/tree/master/vigan/astro}{https://github.com/avigan/Python-utils/tree/master/vigan/astro}} that currently supports 17 different stellar and substellar evolutionary models. Some of these models offer the additional possibility to customize astrophysical parameters such as metallicity, rotational velocity, helium content, alpha enhancement and the spot fraction, for a total of $\sim 110$ different isochrone tracks (Table~\ref{tab:iso_models}. 

\begin{table*}
  \centering
   \caption{Isochrone models currently supported by \textsc{madys} with their bibliographic references. Adopted values for solar metallicity ($Z_\odot$) and the initial helium abundance ($Y_0$) are reported (u=unknown), together with mass ($M$) and age ranges ($t$). Customizable parameters (c.p.): metallicity (m), helium content (Y), alpha enhancement ($\alpha$), rotational velocity (v), spot fraction (s).} 
   \begin{tabular}{lcccccc} \hline \hline
Name & $Y_0$ & $Z_\odot$ & $M$ & $t$ & c. p. & reference \\
 & & & $M_\odot$ & Myr & \\
            \hline
ATMO2020 & 0.275 & 0.0169 & [0.001, 0.075] & $[10^0, 10^4]$ & \textemdash & {\citet{2020A&A...637A..38P}} \\
B97 & 0.27431 & 0.01886 & [$10^{-3}$,0.04] & $[10^0, 10^4]$ & \textemdash & {\citet{1997ApJ...491..856B}} \\
\multirow{2}{*}{BEX} & \multirow{2}{*}{0.27} & \multirow{2}{*}{0.0142} & \multirow{2}{*}{[$1.5 \cdot 10^{-5}$,0.002]} & \multirow{2}{*}{$[1, 4 \cdot 10^3]$} & \multirow{2}{*}{\textemdash} & {\citet{2019A&A...623A..85L};} \\
& & & & & & {\citet{2019A&A...624A..20M}} \\
BHAC15 & 0.271 & 0.0153 & [0.01,1.4] & [0.5, $10^4$] & \textemdash & {\citet{2015A&A...577A..42B}} \\
Geneva & 0.266 & 0.014 & [0.8,120] & [0.1, $10^2$] & \textemdash & {\citet{2019A&A...624A.137H}} \\
MIST & 0.2703 & 0.0142 & [0.1,150] & $[10^{-1},2\cdot 10^4]$ & m, $\alpha$, v & {\citet{2016ApJS..222....8D,2016ApJ...823..102C}} \\
PARSEC & 0.2485 & 0.01524 & [0.09,350] & $[10^{-1}, 10^4]$ & m & {\citet{2017ApJ...835...77M}} \\
(PHOENIX) AMES-COND & 0.247 & 0.018 & [0.005,1.4] & $[1, 1.2 \cdot 10^4]$ &  \textemdash & {\citet{2001ApJ...556..357A}} \\
(PHOENIX) AMES-Dusty & 0.247 & 0.018 & [0.005,1.4] & $[1, 1.2 \cdot 10^4]$ &  \textemdash & {\citet{2001ApJ...556..357A}} \\
(PHOENIX) Bt-Settl & 0.271 & 0.0153 & [0.01,1.4] & $[10^0, 10^4]$ & \textemdash & {\citet{2016sf2a.conf..223A}} \\
(PHOENIX) NextGen & 0.247 & 0.018 & [0.01,1.4] & $[10^0, 1.2 \cdot 10^4]$ & \textemdash & {\citet{1999ApJ...525..871H}} \\
PM13 & \textemdash & \textemdash & [0.075,27] & \textemdash & \textemdash & \citet{2013ApJS..208....9P} \\
SB12 & 0.27431 & 0.01886 & [$10^{-3}$,$10^{-2}$] & [1,100] & \textemdash & \citet{2012ApJ...745..174S} \\
Sonora Bobcat & 0.2735 & 0.0153 & $[5 \cdot 10^{-4},10^{-1}]$ & $[10^{-1}, 10^4]$ & \textemdash & {\citet{2021ApJ...920...85M}} \\
SPOTS & 0.2676 & 0.0165 & [0.1,1.3] & $[10^0, 4 \cdot 10^3]$ & s & {\citet{2020ApJ...891...29S}} \\
STAREVOL & 0.269 & 0.0134 & [0.2,1.5] & $[1, 1.26 \cdot 10^4]$ & m, v & {\citet{2019A&A...631A..77A}} \\
YAPSI & 0.2775 & 0.0142 & [0.15,5] & $[0.5, 10^4]$ & m, Y & {\citet{2017ApJ...838..161S}} \\
  \hline\hline
   \end{tabular}
   \label{tab:iso_models}
\end{table*}

Mass and age ranges for the grid can be externally imposed; otherwise, the program computes suitable mass limits that take into account both the dynamical range of the model and the range of values expected from a rough preliminary evaluation of the sample's absolute Gaia DR3 $G$-band and, if applicable, 2MASS $K_s$-band magnitudes based on the tables by \citet{2013ApJS..208....9P}.

After selecting a model, the program creates two vectors of masses and ages, whose points are equally spaced on a logarithmic scale, and builds the theoretical photometric grid. Linear interpolation between consecutive points into the final grid is performed for every quantity, and no extrapolation is attempted outside the grids. The usage of a fixed grid, rather than real-time interpolation, was chosen for a twofold reason: to handle arbitrarily large group of stars while keeping computation times reasonable, and to allow a homogeneous treatment of statistical uncertainties. The spacing between consecutive steps can be adjusted, but is by default significantly smaller than any realistic uncertainty value.

Generally speaking, the choice of the isochrone set for a given sample should be carefully pondered depending on the expected mass and age ranges of the sample and on the photometric filters of interest. We notice that a few isochrone sets do not come equipped with Gaia and 2MASS filters: hence, they can only be used in mode 2. The program is not halted if it encounters a filter that is not available for the selected model, but --after printing a warning-- it neglects all the photometry provided in that filter. For this reason, it is always recommended to look at the execution log produced by the program.

\subsection{Age and mass determination}
\label{sec:age_mass}

For each object in the final database (Section~\ref{sec:phot_sel}), \textsc{madys} seeks the minimum of a suitable $\chi^2$ function:

\begin{equation}
\chi^2 = \sum_k \left ( \frac{M^{th}_k-M^{obs}_k}{\sigma_{M^{obs}_k}} \right )^2 \equiv \sum_k s^2_k
\end{equation}

which can be thought as a 2D distance matrix with same shape as the age-mass grid and elements:

\begin{equation}
\chi^2_{ij} = \sum_k \left ( \frac{M^{th}_{ijk}-M^{obs}_k}{\sigma_{M^{obs}_k}} \right )^2,
\end{equation}

\noindent where $M^{th}_{ijk}$ is the theoretical magnitude in the $k$-th filter corresponding to the $i$-th mass and $j$-th age of the model grid, $M_{obs,k}$ is the observed magnitude in the same filter and $\sigma_{M_{obs,k}}$ its associated uncertainty. The sum is done only over the filters $k$ passing the following prescriptions:
\begin{enumerate}
    \item an error on the absolute magnitude smaller than 0.2 mag;
    \item a best-match $M^{th}_{i_0 j_0 k}$ such that $|M^{th}_{i_0 j_0 k}-M^{obs}|<0.2$ mag.
\end{enumerate}

Individual age ranges can be provided for each target, and this is particularly useful when external constraints are available; the only caveat is that the kind of input should be the same for every target. In particular, if the age of each object star is explicitly imposed, or a triplet [optimal age, minimum age, maximum age] is provided (case 1), a single filter is sufficient for parameter estimation; conversely, if no age constraint is given, or just a doublet [minimum age, maximum age] is provided (case 2), the estimation is performed only if the following conditions are met:
\begin{enumerate}\addtocounter{enumi}{2}
    \item at least three filters passed the prescriptions 1. and 2.;
    \item after identifying the minimum $\chi^2$, its third smallest associated $s^2_k$ < 9, or alternatively its third smallest $|M^{th}_{k}-M^{obs}_k|<0.1$ mag.
\end{enumerate}

In order to embed photometric uncertainty into the final parameter estimate, the procedure is repeated $q=1000$ times while randomly varying, using a Monte Carlo approach, the apparent photometry and the parallax according to their uncertainties (which are assumed to be distributed in a normal fashion). 

In case 1, the age is not fitted, and the resulting mass distribution is assumed to be unimodal: in other words, the masses corresponding to the $16^{th}$,$50^{th}$ and $84^{th}$ percentile of the sample composed by the $q$ best-fit solutions are returned.

In case 2, the algorithm considers the possibility of a multimodal posterior distribution for both age and mass. At each iteration $q$, having a minimum $\chi^2=\chi^2_q$, the set of $(i,j)$ mass and age steps such that:
\begin{equation}
    \chi^2_{i,j}<\chi^2_q+\Delta \chi^2
\end{equation}
are collected and added to a single array, $\bar{P}$. We decided to adopt $\Delta \chi^2 = 3.3$ as it defines the 68.3\% confidence region around the best-fit joint parameter estimate for a two-parameter model \citep[see, e.g.,][]{2010LNP...800..147V}.

The final outcome of this procedure is an array of solutions, $\bar{P}$. The "hottest points" are the indices recurring more frequently; each occurrence of a point has an associated $\chi^2$, and this should be properly reflected into the final weighted average. In general, the ensemble of points in $\bar{P}$ will not be connected, meaning that multiple families of solutions in the age-mass space can be possible.

An intuitive approach to identify these families consists in identifying connected regions in the age-mass grid. In order to reduce the strong dependence of the connection on random realizations of data perturbations, we decided to define as "attraction points" the points which appear at least in the 10\% of the interactions in $\bar{P}$. Each isolated attraction point defines a family of solutions; a group of contiguous attraction points is treated as a single attraction point located in the group's center of mass, hence defining a single family of solutions as well. The remaining points are then assigned to the family of the closest attraction point.

Each family of solutions $p$ corresponds, from a physical perspective, to a different physical solution; its associated age and mass estimates ($t_p$,$m_p$) are defined as the average of the $i$-th mass and the $j$-th age, weighted by a coefficient $1/\chi^2_{ij,q}$:

\begin{equation}
\log_{10}{m_p}=\frac{\sum_{(i,j) \in p} \log_{10}{m_i} \cdot (\chi^2_{ij,q})^{-1}}{\sum_{(i,j) \in p} (\chi^2_{ij,q})^{-1}},
\end{equation}

\begin{equation}
\log_{10}{a_p}=\frac{\sum_{(i,j) \in p} \log_{10}{a_j} \cdot (\chi^2_{ij,q})^{-1}}{\sum_{(i,j) \in p} (\chi^2_{ij,q})^{-1}},
\end{equation}

\noindent where, of course, points $(i,j)$ repeating in different iterations are summed each time with a weight corresponding to the $\chi^2_{ij,q}$ of the $q$-th iteration.

The variances associated to $\log_{10}{m_p}$ and $\log_{10}{a_p}$ are given by:
\begin{equation}
\sigma^2_{m_p} = \frac{\sum_{(i,j) \in p} (\log_{10}{m_i}-\log_{10}{m_p})^2 \cdot (\chi^2_{ij,q})^{-1}}{\sum_{(i,j) \in p} (\chi^2_{ij,q})^{-1}},
\end{equation}
\begin{equation}
\sigma^2_{a_p} = \frac{\sum_{(i,j) \in p} (\log_{10}{a_i}-\log_{10}{a_p})^2 \cdot (\chi^2_{ij,q})^{-1}}{\sum_{(i,j) \in p} (\chi^2_{ij,q})^{-1}}.
\end{equation}

Couples of solutions ($p_1$,$p_2$) that are consistent with representing the same solution, that is to say with:
\begin{equation}
\Delta d=\frac{(\log_{10}{m_{p_1}}-\log_{10}{m_{p_2}})^2}{\sigma^2_{m_{p_1}}+\sigma^2_{m_{p_2}}} + \frac{(\log_{10}{a_{p_1}}-\log_{10}{a_{p_2}})^2}{\sigma^2_{a_{p_1}}+\sigma^2_{a_{p_2}}} < 8,
\end{equation}
are merged together. The outcome of the process is a set of solutions $\{p\}$, each one bearing a fraction of the total region of the solutions $\bar{P}$ equal to:
\begin{equation}
w_p = \frac{\sum_{(i,j) \in p} (\chi^2_{ij,q})^{-1}}{\sum_{(i,j) \in \bar{P}} (\chi^2_{ij,q})^{-1}}.
\end{equation}
The solution with the maximum $w_p$ is returned as the best-fit solution, but the other solutions can be inspected as well. Both the $\chi^2$ map for nominal photometric values and the weight map $W$, defined as the 2D matrix with elements:
\begin{equation}
w_{ij}=\frac{1}{\sum_{(i,j) \in \bar{P}} (\chi^2_{ij,q})^{-1}}
\end{equation}
referring instead to the whole fitting process, can be returned and plotted through a dedicated method.

\section{Applications}
\label{sec:applications}

\subsection{Age substructures}
\label{sec:scocen}

The ability of \textsc{madys} to handle thousands of stars at once makes it particularly suited to the systematic study of young ($t \gtrsim 5$ Myr) regions with a clear kinematic fingerprint. Indeed, the requirement of a significant portion of the stellar sample in the pre-MS phase and the caveat for the derived extinctions (Section~\ref{sec:ext}) naturally define an optimal age range for \textsc{madys} between 5-10 Myr and a few hundred million years.

As a possible application of the code, we compute here the age of confirmed members of the Scorpius-Centaurus association. The association, that is the nearest star-forming region to the Sun, is classically divided into three subgroups: Upper Scorpius (US), Upper Centaurus-Lupus (UCL) and Lower Centaurus-Crux (LCC) \citep{1999AJ....117..354D}.

We start from the list of bona fide Scorpius-Centaurus members compiled by \citet{2019A&A...623A.112D} using Gaia DR2 data. In order to define the subregions, we employ classical coordinate boundaries as in \citet{1999AJ....117..354D} and subsequent works: for US, $l \in [343^\circ, 360^\circ]$, $b \in [10^\circ,30^\circ]$; for UCL, $l \in [313^\circ, 343^\circ]$, $b \in [0^\circ,25^\circ]$; for LCC, $l \in [280^\circ, 313^\circ]$, $b \in [-10^\circ,23^\circ]$.

Starting from Gaia DR2 IDs, \textsc{madys} recovers the photometry and computes extinction values as described in Section~\ref{sec:ext}. The age and mass determination, initialized with only a modest constraint on age ($t \in [1,300]$ Myr), is done here with the BHAC15 models \citep{2015A&A...577A..42B}.

A visual inspection of the ($G_{BP}-G_{RP}, G$) CMD shows that some stars appear to be too old to be members of the association, and indeed they have fitted ages $\gtrsim 100$ Myr. Therefore, we exclude the stars with best-fit ages greater than 60 Myr; for the few stars with multiple possible solutions, meaning that there is an overlap between a pre-MS solution and an evolved MS solution, we pick the youngest one.

The derived ages and masses for the three subgroups, computed as the $16^{th}$,$50^{th}$ and $84^{th}$ percentile of the age distribution of their members, are:

\begin{equation*}
    US: 6.6_{-3.0}^{+5.6} \text{ Myr,}
\end{equation*}
\vspace{-5mm}
\begin{equation*}
    UCL: 9.1_{-3.7}^{+4.8} \text{ Myr,}
\end{equation*}
\vspace{-5mm}
\begin{equation*}
    LCC: 8.7_{-3.2}^{+5.6} \text{ Myr.}
\end{equation*}

\begin{figure}
	\includegraphics[width=\hsize]{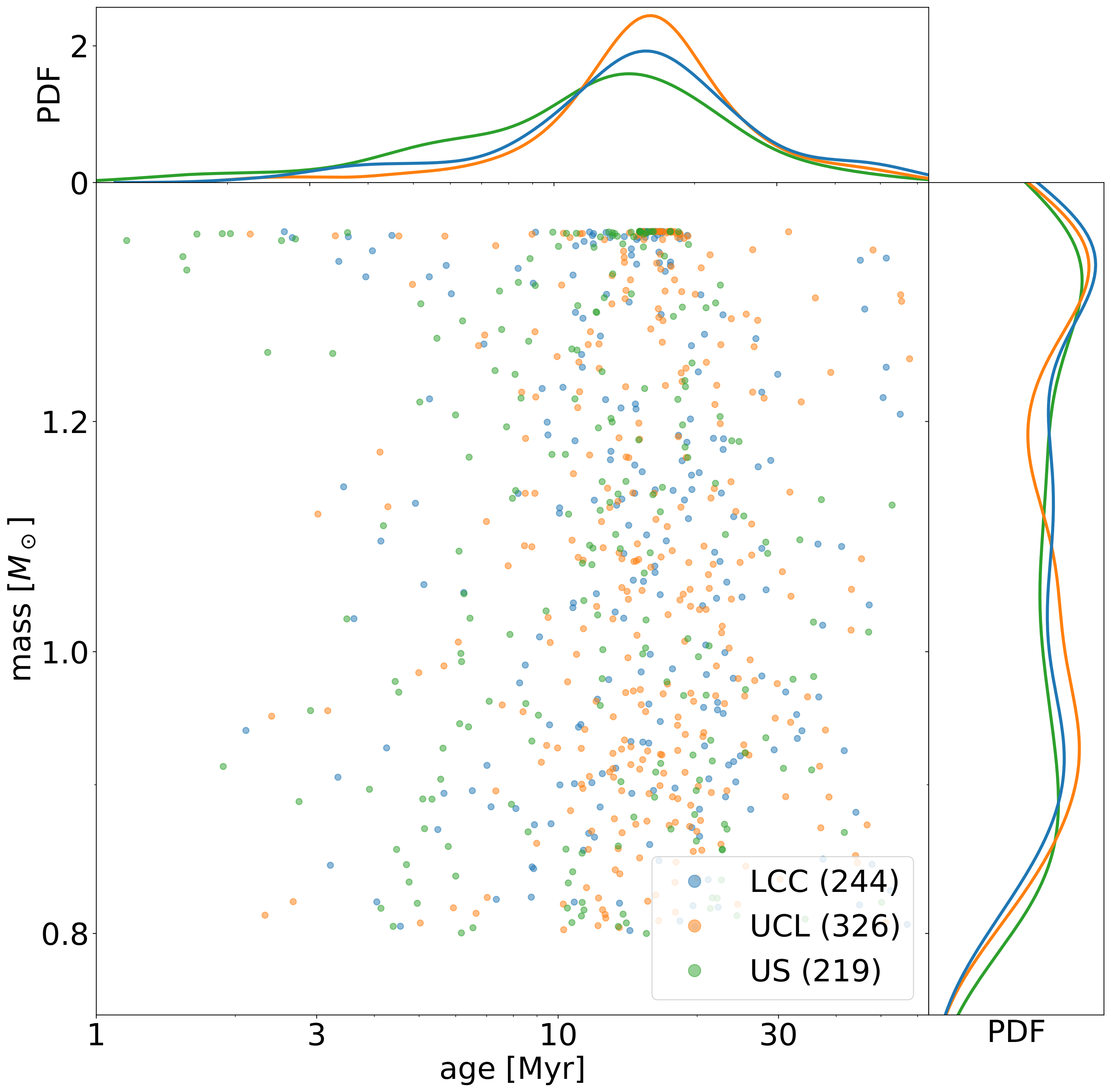}
    \caption{Derived age and mass distribution of the Sco-Cen sample, restricted to stars with $M>0.8 M_\odot$. The insets represent a kernel-density estimate of the underlying distributions. The clump of sources at the upper mass end is an artifact due to the sharp mass cut of BHAC15 isochrones at $1.4 M_\odot$.}
    \label{fig:density_plot}
\end{figure}

We recover some facts which are already known from the literature \citep[see, e.g.][]{2016MNRAS.461..794P}: firstly, the observation that US is younger than UCL and LCC; secondly, the existence of a positive correlation between age and mass; in other words, M stars appear younger than their F and G counterparts \citep[for a review, see][]{2021MNRAS.507.1381S}. Although an age spread between the two cannot be completely ruled out, most of the observed spread is likely due to a slowed-down contraction of low-mass stars caused by magnetic fields \citep{2016A&A...593A..99F}. Indeed, if we restrict to stars with a best-fit $M>0.8 M_\odot$ (Figure~\ref{fig:density_plot}), the results become: $US: 13.1_{-7.1}^{+8.3} \text{ Myr}$, $UCL: 16.0_{-5.0}^{+6.9} \text{ Myr}$, $LCC: 15.3_{-6.4}^{+7.9} \text{ Myr}$. The results are similar, both in the median value and in the associated scatter, to the estimates by \citet{2016MNRAS.461..794P}.

With these caveats in mind, the possibility of computing individual age estimates for pre-MS stars with \textsc{madys} opens up important opportunities for the study of young star-forming regions, whose exquisite substructures are being more and more connected with their star formation history \citep[e.g.,][]{2021ApJ...917...23K,2021AJ....162..110K,2021MNRAS.507.1381S}.

\subsection{Stellar physical parameters}
Although by construction \textsc{madys} is able to return age estimates for main-sequence stars, we stress that in this case they should be regarded as purely indicative. Nevertheless, the argument can be reversed: if external age constraints are available, \textsc{madys} can return precise determination of stellar parameters such as mass, effective temperature, radius and surface gravity for large samples of stars.

As an example of this possibility, we recovered from the literature a collection of stars with interferometrically measured angular diameters. Our sample combines the six main-sequence stars studied by \citet{2012ApJ...760...32H}
and the full samples by \citet{2012ApJ...746..101B,2012ApJ...757..112B}, spanning a spectral range that stretches from A- to M-type. We combined angular measurements with the latest parallaxes from Gaia DR3 to have radius estimates with median precision $\sim 1\%$.

Our parameter estimates are based on PARSEC isochrones; we applied only a modest age constraint ($t \in [500, 7000]$ Myr); with respect to metallicity, we refer to [Fe/H] estimates from the original studies. Under the assumption [Fe/H]$\approx$[M/H], we used for each star the isochrone grid with the closest metallicity rather than interpolating \footnote{The available metallicities for this example were: [Fe/H]=[-1.0,-0.75,-0.5,-0.25,0.0,0.13,0.25,0.5,0.75,1.00].}.

\begin{figure}
	\includegraphics[width=\hsize]{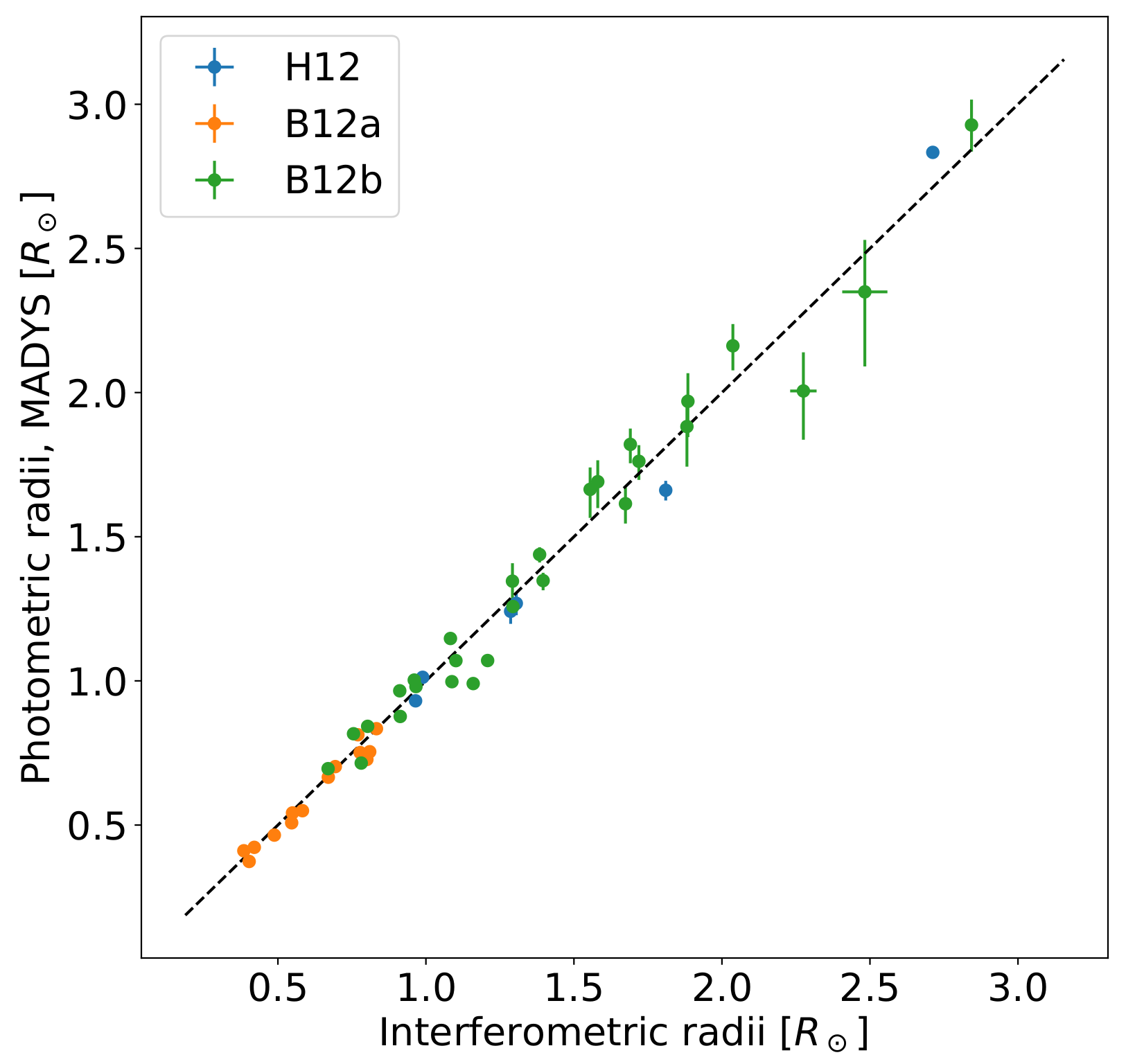}
    \caption{Comparison between photometric radius estimates obtained by \textsc{madys} and interferometric radii from the literature: \citet{2012ApJ...760...32H} (H12), \citet{2012ApJ...746..101B} (B12a), \citet{2012ApJ...757..112B} (B12b).}
    \label{fig:radii}
\end{figure}

The results are shown in Figure~\ref{fig:radii}. The mean and standard deviation of the fractional difference between interferometric radii and those returned by \textsc{madys} are +1\% and 6\%, respectively.

\subsection{Mass of directly imaged substellar companions}
\label{sec:planets}

\textsc{madys} has been conceived to connect models spanning from young self-luminous gas giants to old stars. Thanks to the complete coverage of JWST filters offered by some models, \textsc{madys} will be a valuable tool to study the first data coming from the instrument, and to help characterize the properties of newly discovered objects.

Indeed, mode 2 is intended for objects that are not found in Gaia, such as objects discovered in direct imaging studies, either in isolation or as companions to stellar objects. In the latter case, \textsc{madys} can be used in two steps of the chain: to obtain the age of the stellar target --either directly or by exploiting its kinematic neighbors-- and to use such information to derive a mass estimate for the companion. A combination of indirect kinematic constraints and literature data was recently applied in \citet{2022A&A...664A...9S} to derive age and mass estimates for both the stellar target and its companions.

We present here an application of \textsc{madys} to the HR 8799 system \citep{2008Sci...322.1348M}, one of the cornerstones of direct imaging studies. With four giant planets detected around a $1.47^{+0.11}_{-0.08} M_\odot$ primary \citep{2018AJ....156..192W}, the system is a unique laboratory to test the accuracy of substellar evolutionary models.

Several age estimates have been derived for the system in the recent literature: $10-23$ Myr from \cite{2022AJ....163...52S}, $33^{+7}_{-13}$ Myr from \cite{2012ApJ...761...57B} or $42^{+6}_{-4}$ Myr from \cite{2015MNRAS.454..593B}; we notice that the last estimate, based on the lithium-depletion boundary for the Colomba association which the star appears to be a member of, is independently indicated by \textsc{madys} when inspecting the kinematic neighborhood of the star. Indeed, we identified three stars\footnote{Gaia EDR3 2838213864935858816, Gaia EDR3 2835796794780262912 and Gaia EDR3 2830197806693622272.} with projected separation < 3 pc and tangential velocity difference < 3 km s$^{-1}$: all of them have a best-fit mass $\in [0.7, 1] M_\odot$ and age $\sim 40$ Myr.

Nevertheless, we conservatively start from two possible age intervals, $t \in [10, 23]$ Myr and $t \in [30, 60]$ Myr, to compare our estimates with already published results. As a consequence of the uncertainty on age, we expect the model-dependent uncertainty on the derived photometric mass estimates to be broadened.

Table~\ref{tab:hr8799} reports literature estimates for the masses of the four planets, obtained with photometric or dynamical methods, together with new estimates obtained by \textsc{madys}. We collected contrasts measurements from \citet{2016A&A...587A..57Z} in the Spectro-Polarimetric High-contrast Exoplanet Research \citep[SPHERE;][]{2019A&A...631A.155B} bands $J$ ($\lambda_{peak}=1.245~\mu$m), $H_2$ ($\lambda_{peak}=1.593~\mu$m), $H_3$ ($\lambda_{peak}=1.667~\mu$m), $K_1$ ($\lambda_{peak}=2.110~\mu$m) and $K_2$
($\lambda_{peak}=2.251~\mu$m), and combined them to 2MASS magnitudes and Gaia DR3 parallax for HR 8799 to obtain absolute magnitudes.

Mass estimates were obtained through four models: namely, AMES-Dusty, AMES-Cond, ATMO2020 (in particular, the chemical equilibrium grid) and Sonora Bobcat \footnote{For ATMO2020 and Sonora Bobcat, which currently lack SPHERE filters, we employed theoretical magnitudes in the closest photometric system available: the Mauna Kea Observatories photometric system \citep[MKO, ][]{2002PASP..114..180T}, and 2MASS, respectively. In particular, $J_{\text{SPHERE}} \sim J_{\text{MKO}} \sim J_{\text{2MASS}}$, $K_{\text{SPHERE}} \sim K_{\text{MKO}} \sim K_{s,\text{2MASS}}$, $0.5 \cdot (H_{2,\text{SPHERE}}+H_{3,\text{SPHERE}}) \sim H_{\text{MKO}} \sim H_{\text{2MASS}}$.}.

\begin{figure*}
	\includegraphics[width=\hsize]{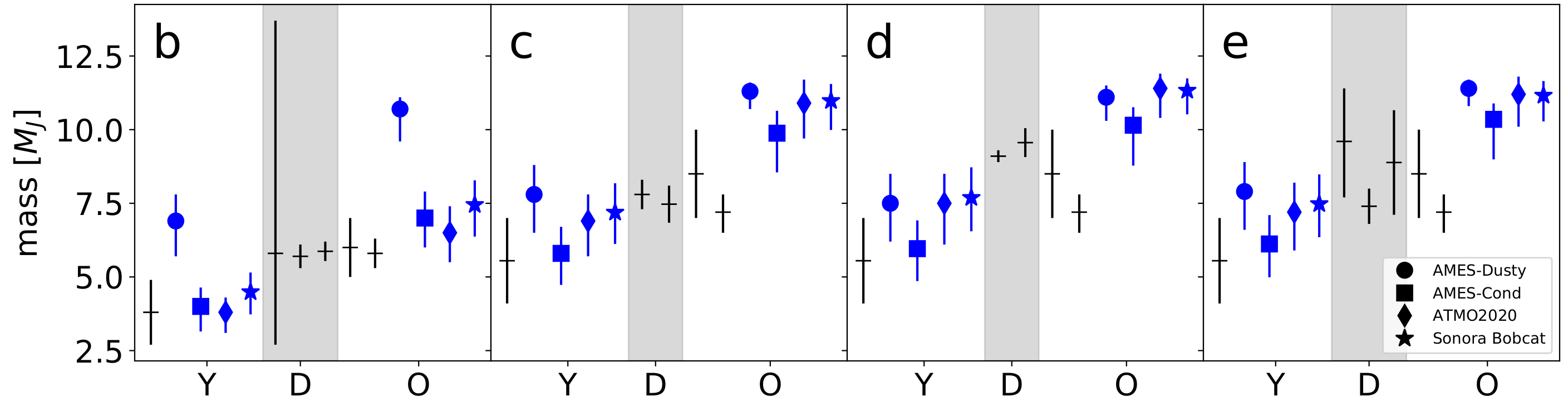}
    \caption{Literature (black) and new (blue) mass estimates for the HR 8799 planets. Each panel refers to the eponymous planet. For the sake of clarity, dynamical estimates (D) have been placed in the gray region, which visually separates the estimates based on a younger (10-23 Myr) age (Y) from those estimates based on an older (30-60 Myr) age (O). Different symbols for \textsc{madys} estimates refer to different models.}
    \label{fig:HR8799}
\end{figure*}

The results are also summarized in Figure~\ref{fig:HR8799}. While the results of photometric estimates can significantly differ from one another even in the same age window, tighter dynamical constraints\footnote{The small errorbar of the dynamical mass estimates by \citet{2020ApJ...902L..40G} is a consequence of the assumption that the planets are in an exact 8:4:2:1 mean-motion resonance, and should therefore be taken with caution.} coming from thorough astrometric follow-up in the next few years will help distinguishing among them, shedding light into the still poorly constrained cooling timescale of young self-luminous Super Jupiters.

\begin{table*}
  \centering
   \caption{Mass estimates for the planets in the HR 8799 system. We compare the best-fit results from \textsc{madys} with photometric and dynamical masses taken from the literature. Two age ranges ($t \in 10-23$ Myr, $t \in 30-60$ Myr) and four models are used for the independent estimates.}
   \begin{tabular}{c|c|c|c|c|c|c} \hline
source & age & b & c & d & e & method \\
 & Myr & \mj & \mj & \mj & \mj & \\
            \hline
{\citet{2008Sci...322.1348M}} & $30-60$ & $5-7$ & $7-10$ & $7-10$ & $7-10$ & photometric \\[1.4mm]
{\citet{2018ApJ...855...56W}} & \textemdash & $5.8^{+7.9}_{-3.1}$ & \textemdash & \textemdash & \textemdash & dynamical \\[1.4mm]
{\citet{2018AJ....156..192W}} & $42 \pm 5$ & $5.8 \pm 0.5$ & $7.2_{-0.7}^{+0.6}$ & $7.2_{-0.7}^{+0.6}$ & $7.2_{-0.7}^{+0.6}$ & photometric \\[1.4mm]
{\citet{2020ApJ...902L..40G}} & & $5.7 \pm 0.4$ & $7.8 \pm 0.5$ & $9.1 \pm 0.2$  & $7.4 \pm 0.6$ & dynamical \\[1.4mm]
{\citet{2021ApJ...915L..16B}} & $42^{+24}_{-16}$ & \textemdash & \textemdash & \textemdash & $9.6^{+1.8}_{-1.9}$ & dynamical \\[1.4mm]
{\citet{2022AJ....163...52S}} & $10-23$ &  $2.7-4.9$ & $4.1-7.0$ & $4.1-7.0$ & $4.1-7.0$ & photometric \\[1.4mm]
{\citet{zurlo}} & \textemdash & $5.54-6.20$ & $6.84-8.10$ & $9.07-10.05$ & $7.11-10.66$ & dynamical \\[1.4mm] \hline
\multirow{8}{*}[-1.5em]{\textsc{madys} (this work)} & $10-23$ &  $6.9_{-1.2}^{+0.9}$ & $7.8_{-1.3}^{+1.0}$ & $7.5_{-1.3}^{+1.0}$ & $7.9_{-1.3}^{+1.0}$ & AMES-Dusty \\[1.4mm]
 & $30-60$ & $10.7_{-1.1}^{+0.4}$ & $11.3_{-0.6}^{+0.3}$ & $11.1_{-0.8}^{+0.4}$ & $11.4_{-0.6}^{+0.3}$ & AMES-Dusty \\[1.4mm]
 & $10-23$ & $3.8_{-0.7}^{+0.5}$ & $6.9_{-1.2}^{+0.9}$ & $7.5_{-1.4}^{+1.0}$ & $7.2_{-1.3}^{+1.0}$ & ATMO2020 \\[1.4mm]
 & $30-60$ & $6.5_{-1.0}^{+0.9}$ & $10.9_{-1.2}^{+0.8}$ & $11.4_{-1.0}^{+0.5}$ & $11.2_{-1.1}^{+0.6}$ & ATMO2020 \\[1.4mm]
 & $10-23$ & $4.0_{-0.8}^{+0.6}$ & $5.8_{-1.1}^{+0.9}$ & $6.0_{-1.1}^{+1.0}$ & $6.1_{-1.1}^{+1.0}$ & Ames-Cond \\[1.4mm]
 & $30-60$ & $7.0_{-1.0}^{+0.9}$ & $9.9_{-1.3}^{+0.8}$ & $10.1_{-1.4}^{+0.6}$ & $10.4_{-1.4}^{+0.5}$ & Ames-Cond \\[1.4mm]
 & $10-23$ & $4.5_{-0.8}^{+0.7}$ & $7.2_{-1.1}^{+1.0}$ & $7.7_{-1.1}^{+1.0}$ & $7.5_{-1.1}^{+1.0}$ & Sonora Bobcat \\[1.4mm]
 & $30-60$ & $7.5_{-1.1}^{+0.8}$ & $11.0_{-1.0}^{+0.6}$ & $11.3_{-0.8}^{+0.4}$ & $11.2_{-0.9}^{+0.5}$ & Sonora Bobcat \\
\hline
   \end{tabular}
   \label{tab:hr8799}
\end{table*}
\noindent

\section{Discussion}
\label{sec:discussion}

Since the algorithm behind age determination in \textsc{madys} is based on isochronal fitting, the tool automatically inherits the same drawbacks of this technique, which have been the subject of extensive discussion in the literature \citep[see, e.g.,][]{2010ARA&A..48..581S,2016EAS....80..115B}.

In particular, \textsc{madys} alone cannot be used to understand if a star is young ($t \lesssim 100$ Myr) or not: a degeneracy exists between young pre-MS stars and evolved stars that have left the MS, and this is naturally reflected into different families of solutions that arise if the age is left completely unconstrained. A young solution is to be preferred if independent youth indicators (e.g., activity indicators such as X-ray, UV, $H_\alpha$ emission) are available. 

A conceptually different youth indicator is the membership to a young star-forming region. Indeed, the integration of kinematic information into \textsc{madys} will be the subject of a second version of the tool. For the moment being, \textsc{madys} can exploit existing lists of confirmed members of these regions to unveil correlations between the star formation history and kinematic substructures \citep{2021MNRAS.507.1381S}.

A strong asset of \textsc{madys} is the ability to collect and handle photometric data for thousands of stars. The age determination is rapid, taking about one second per star under typical conditions. In this sense, our tool constitutes a step forward in the automation of the process with respect to existing tools such as \texttt{PARAM} \citep{2017MNRAS.467.1433R}, \texttt{ARIADNE} \citep{2022MNRAS.tmp..920V}, \texttt{stardate} \citep{2019JOSS....4.1469A} or \texttt{isochrones} \citep{2015ascl.soft03010M}; on the other hand, unlike them, it does not currently allow one to exploit additional information coming, for instance, from asteroseismology or spectroscopy during the fitting process.

A second strength of \textsc{madys} is the possibility to employ a large collection of stellar and substellar models, allowing the possibility to evaluate the impact of different input physics into the final results. This is particularly important not only for the pre-MS phase of stars, but also in the context of direct imaging studies of cooling substellar objects, where there is still no established standard on how photometric mass determinations are to be obtained.

\section{Conclusions}
\label{sec:conclusions}

We introduced a Python tool, \textsc{madys}, aimed at obtaining photometric age and mass estimates for arbitrarily large groups of young stellar or substellar objects. The main strengths of the tool are:

   \begin{itemize}
      \item the ability to query and crossmatch different catalogs to yield highly reliable catalogs of potentially large lists of objects;
      \item the possibility to take interstellar extinction into account;
      \item the ability to derive photometric ages and mass estimates by comparing dozens of filters with a large suite of substellar or stellar evolutionary models;
      \item the possibility to unambiguously compare the predictions of different models, and to see the effect of changing astrophysical parameters;
      \item the large plotting options for efficient data visualization.
   \end{itemize}

These features give \textsc{madys} a large number of possible scientific applications, such as:
\begin{itemize}
    \item the study of young star-forming regions, connecting kinematic data with age determinations;
    \item direct-imaging studies, including forthcoming JWST observations. Even in the case of a nondetection, the tool can be useful to turn contrast limit into mass limits, paving the way to a systematic assessment of the demographics of direct-imaged exoplanets and brown dwarfs.
\end{itemize}

\noindent Besides the inclusion of new models and filters, future developments of \textsc{madys} will include the possibility to do the following: simultaneously deriving extinction and (sub)stellar parameters under the current optimization scheme; implementing an indirect method for age determination based on empirical kinematic properties; and, finally, providing a systematic comparison of isochronal and kinematic results with those obtained through other age determination techniques.

\begin{acknowledgements}
We are grateful to Dr. N{\'u}ria Miret-Roig, Dr. Raffaele Gratton and Dr. Gabriel-Dominique Marleau for their precious advice. We are extremely grateful to the anonymous referee for the constructive comments, which significantly helped raise the quality of this paper.

This research has made use of the VizieR catalogue access tool, CDS, Strasbourg, France (DOI : 10.26093/cds/vizier). The original description of the VizieR service was published in 2000, A\&AS 143, 23. This research has made use of the SIMBAD database, operated at CDS, Strasbourg, France.

This work has made use of data from the European Space Agency (ESA) mission {\it Gaia} (\url{http://www.cosmos.esa.int/gaia}), processed by the {\it Gaia} Data Processing and Analysis Consortium (DPAC, \url{http://www.cosmos.esa.int/web/gaia/dpac/consortium}). Funding for the DPAC has been provided by national institutions, in particular the institutions participating in the {\it Gaia} Multilateral Agreement. 
This publication makes use of data products from the Two Micron All Sky Survey, which is a joint project of the University of Massachusetts and the Infrared Processing and Analysis Center/California Institute of Technology, funded by the National Aeronautics and Space Administration and the National Science Foundation. 
VS acknowledges the support of PRIN-INAF 2019 Planetary Systems At Early Ages (PLATEA).

For the purpose of open access, the authors have agreed to apply a Creative Commons Attribution (CC BY) licence to any Author Accepted Manuscript version arising from this submission.

\end{acknowledgements}

%
%

\bibliographystyle{aa}
\bibliography{biblio}

%

\end{document}